\documentclass[aps, pra, reprint]{revtex4-2}
\usepackage[utf8]{inputenc}
\usepackage[T1]{fontenc}
\usepackage{amsmath}
\usepackage{amssymb}
\usepackage{graphicx}
\usepackage{physics}
\usepackage{dsfont}
\usepackage{natbib}
\usepackage{multirow}
\usepackage{xparse} % for \NewDocumentCommand
\usepackage{acronym}
\usepackage{hyperref}
\usepackage{subfigure}
\usepackage{xcolor}

\newcommand{\etaNM}[2]{\eta\left(N, M\right)}

\newcommand{\ExpO}[2]{\mathop{\mathds{E}}_{#2}\left[#1\right]}

\newcommand{\sNorm}[1]{\left| #1 \right|}
\NewDocumentCommand{\QC}{ m o }{%
	\mathcal{#1}%
	\IfNoValueF {#2} {\left( #2 \right)}%
}

\newacro{QKD}{quantum key distribution}
\newacro{UQCM}{universal quantum cloning machine}
\newacro{SQC}{symmetric quantum cloning}
\newacro{AQC}{asymmetric quantum cloning}
\newacro{SQC}{symmetric quantum cloning}
\newacro{CV}{continuous variable}
\newacro{MIMO}{multiple-input multiple-output}
\newacro{Tx}{transmitter}
\newacro{Rx}{receiver}
\newacro{CSI}{channel state information}
\newacro{DMT}{diversity multiplexing tradeoff}
\newacro{RF}{radio frequency}
\newacro{FSO}{free-space optical}

\begin{document}
\title{Diversity and Multiplexing in Quantum MIMO Channels}
\author{Junaid ur Rehman}
\email{junaid.urrehman@uni.lu}

\author{Leonardo Oleynik}

\author{Seid Koudia}

\author{Mert Bayraktar}

\author{Symeon Chatzinotas}
\affiliation{Interdisciplinary Centre for Security, Reliability, and Trust (SnT), University of Luxembourg, L-1855 Luxembourg City, Luxembourg.}

\begin{abstract}
Characterization and exploitation of multiple channels between the transmitter and the receiver in multiple-input multiple-output (MIMO) communications brought a paradigm shift in classical communication systems. The techniques developed around MIMO communication systems not only brought unprecedented advancements in the communication rates but also substantially improved the reliability of communication, measured by low error rates. Here, we explore the same ideas in the paradigm of quantum MIMO communication. Specifically, we utilize approximate quantum cloning to transmit multiple copies of the same quantum state over a MIMO channel that incorporates crosstalk, losses, and depolarizing noise. With this strategy, we find an achievable tradeoff between the average fidelity and communication rate over this MIMO setup.
\end{abstract}
\maketitle
\section{Introduction}
\Ac{MIMO} communication systems brought a revolution in the field of classical communications, specially for the advanced digital communication, such as cellular, Wi-Fi, and digital subscriber line (DSL) communications. Exploiting multiple channels makes it possible to enhance the communication rate by sending multiple data streams in parallel and/or improve the communication reliability by transmitting multiple copies of the same data stream in parallel. The first type of performance improvement is called the multiplexing gain and the second type is called the diversity gain. The main complication, when using multiple channels over the same degree of freedom, is that energy leaks among the channels leading to effects known as radio interference or crosstalk. Diversity schemes enhance the error performance of communication systems by leveraging the statistically different signal degradation experienced across various transmission paths \cite{Wit:91:conf, SW:93:conf, Ala:98:JSAC}. Interestingly, diversity and multiplexing gains in \ac{MIMO} communication are not independent of each other but exhibit a tradeoff relationship known as the \ac{DMT} \cite{ZT:03:TIT}.
Soon after the success of \ac{MIMO} communication techniques in \ac{RF} wireless communication, similar schemes for \ac{FSO} communications were developed and achievable diversity and multiplexing gains were analyzed \cite{SC:02:conf, RS:05:TWC, JB:19:TIT}. Here, a \ac{MIMO}-\ac{FSO} system is equipped with $N_{\mathrm{t}}$ lasers/LEDs and $N_{\mathrm{r}}$ photo-detectors or a multi-aperture telescope with $N_{\mathrm{r}}$ apertures.	

Quantum communication systems equipped with multiple transmitters and/or receivers may benefit from similar reasoning. Situations concerning combinations of quantum channels are usually more interesting than those of combining classical communication channels \cite{KCS:22:CST}. Here we are interested in designing a quantum communication strategy over \ac{MIMO} channels to obtain similar diversity and multiplexing gains. However, there are two fundamental challenges in designing similar diversity schemes in the quantum regime. First, if the information to be transmitted is truly quantum, it is not possible to transmit multiple perfect copies of it. This is due to the fundamental restriction on cloning of arbitrary quantum information \cite{WZ:82:Nat}. This restriction limits the transmitter diversity to be possible only in select few cases, e.g., when the information to be transmitted is classical or the transmitter can exactly implement the source of quantum information. Second, the methods to combine general quantum states received at multiple apertures of a receiving telescope are not yet defined. Similar challenges appear in designing effective techniques for multiplexing gain in quantum communications. For example, one common approach for multiplexing gain in wireless communications is to transmit a classical superposition (namely, weighted sum of complex numbers) of two data streams and successively separating them at the receiver through joint processing of both received signals. In quantum communications, another no-go theorem forbids the creation of superposition of unknown arbitrary quantum states \cite{Som:20:PRA, OGH:16:PRL}.

Due to these fundamental challenges, there have been very few works attempting to design diversity or spatial multiplexing schemes in quantum communications. One notable work \cite{YC:20:TCOM} introduced receiver diversity in FSO quantum communication in turbulent channels. The considered communication system is based on coherent states for binary phase-shift keying. Another recent work exploited spatial diversity in Earth-to-satellite quantum communication links and demonstrated improved communication reliability for entanglement distribution and the transfer of coherent states \cite{WRA:24:arX}. Other works dealing with FSO quantum MIMO communication systems (a) restrict to the case where the information is classical, i.e., quantum key distribution, and (b) utilize multiple antenna systems only for the multiplexing gain instead of achieving the diversity gain \cite{KDM:21:CL, KMC:23:TQE, ZPD:23:TQE}. Therefore, existing approaches fail to achieve a general spatial diversity gain in truly quantum communication systems.

In this work, we investigate diversity and multiplexing for quantum communications on \ac{MIMO} quantum channels. We first begin by developing a basic $2\times 2$  quantum \ac{MIMO} model that include crosstalk between channels, losses, and noise. Then, we develop a diversity achieving scheme in this $2\times 2$ setup. We consider the average (over data streams and time) fidelity of communication and the average (over data streams and time) communication rate (qubit/s) as the figures-of-merit. We demonstrate a clear improvement in the average fidelity when the proposed diversity scheme is employed.  Later, we generalize this model to $2^m \times 2^m$ \ac{MIMO} setup by recursively utilizing out $2\times 2$ model and obtain achievable curves of quantum \ac{DMT} with our proposed scheme. 

\section{Diversity and Multiplexing in MIMO Quantum Communications}
In this section, we present develop our first example of diversity and multiplexing for quantum communications. We begin with our system model of $2 \times 2$ \ac{MIMO} quantum channel, followed by a simple communication strategy that amounts to multiplexing on this \ac{MIMO} setup. This is followed by a subsection on diversity achieving scheme by utilizing \ac{SQC} at the \ac{Tx}.

\subsection{System Model}
\begin{figure}
	\centering
	\subfigure[~]{
	\includegraphics[width=0.475\textwidth]{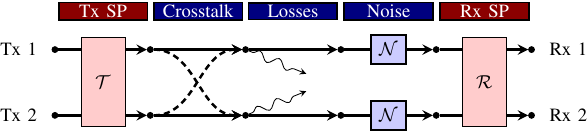}
	}
~~
	\subfigure[~]{
	\includegraphics[width=0.3\textwidth]{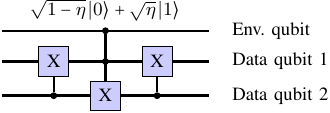}
	}
	\caption{A $2\times 2$ \ac{MIMO} setup with two \acp{Tx} and two \acp{Rx}. (a) A general \ac{MIMO} system with a $2\times 2$ setup including crosstalk, losses, and noise.  The losses and noise are modeled by the erasure channel with erasure probability $\epsilon$ and the depolarizing noise of depolarizing strength $\lambda$, respectively. The crosstalk is parameterized by the crosstalk strength $\eta$ and is modeled by a control SWAP gate, as shown in (b).} 
	\label{fig:diversitycrosstalk}
\end{figure}

We consider a $2\times 2$ quantum \ac{MIMO} communication channel between the \ac{Tx} and the \ac{Rx} as shown in the Fig.~\ref{fig:diversitycrosstalk}(a). The first effect in the channel that we consider is the crosstalk, parameterized by the parameter $\eta \in \left[0, 1\right]$. This is modelled by a control-SWAP gate, also known as the Fredkin gate. The control qubit is part of the environment and is assumed to be in the state $\ket{c}_{\mathrm{env}} = \sqrt{1 - \eta}\ket{0} + \sqrt{\eta}\ket{1}$, as shown in Fig.~\ref{fig:diversitycrosstalk}(b). The crosstalk is followed by channel losses, modelled by independent qubit erasure channels of equal erasure probability $\epsilon$ on each communication link:
\begin{equation}
	\QC{E}_{\epsilon}\left(\rho\right) = \left(1 - \epsilon\right)\rho + \epsilon\ketbra{\epsilon}{\epsilon},
\end{equation} 
where $\ket{\epsilon}$ is the erasure signal orthogonal to the input Hilbert space, i.e., $\sNorm{\bra{\epsilon}\ket{\phi}} = 0$, where $\ket{\phi}$ is an arbitrary state in the input Hilbert space. Finally, the noise is modeled by independent depolarizing channels of equal strengths $\lambda$ on each communication link:
\begin{equation}
	\QC{D}_{\lambda}\left(\rho\right) = \left(1 - \lambda\right)\rho + \lambda\pi,
\end{equation} 
where $\pi = I/2$ is the maximally mixed state. This noise models both the over-the-air impairments and the measurement noise at the receiver. The depolarizing noise acts only if no erasure occurred in the erasure channel. Thus, we can denote our \ac{MIMO} channel with a triple $\left(\eta, \epsilon, \lambda\right)$, where the exact values of these parameters are not known at the \ac{Tx}, but may be known at the \ac{Rx}, as is the standard assumption in the communications literature. We consider the situation where the parameters are perfectly known at the receiver. For more details on channel characterization, the readers are referred to \cite{KOB:24:arXiv}.

\subsection{Communication Strategies}
In this subsection, we detail the two communication strategies that we employ for transmitting quantum states between the transmitter and the receiver. 

\subsubsection{Multiplexing}
In multiplexing, each of the transmitter $i \in \left\{1, 2\right\}$ prepares its own quantum state to be transmitted to the receiver $i$. Since the \ac{CSI} is known at the receivers, they can swap their received states if $\eta > 0.5$. Let $\ket{\psi}_{1, 2} = \ket{\psi}_1 \otimes \ket{\psi}_2$ be the bipartite input where the subsystem $i\in \left\{1, 2\right\}$ is input to the \ac{Tx} $i$. The inputs $\ket{\psi}_1$ and $\ket{\psi}_2$ are assumed to be independent Haar random states. Then, the output $\rho_{1, 2}$ at the \ac{Rx} is:
\begin{align}
	\rho_{1, 2} =\QC{D}_{\lambda}\otimes\QC{D}_{\lambda}\left(
						\QC{E}_{\epsilon}\otimes\QC{E}_{\epsilon}\left( 
						\QC{C}_{\eta}\left(
							\ketbra{\psi}{\psi}_{1, 2}
						\right)
						\right)	
	\right),
\end{align}
where $\QC{C}_{\eta}\left(\sigma \otimes \omega\right) = \left(1 - \eta\right)\sigma \otimes \omega + \eta\, \omega\otimes\sigma$ is the crosstalk noise operating jointly on the two inputs. 

%%%%%%%%%%%%%
\begin{figure*}[ht!]
	\centering
	\includegraphics[width=0.95\textwidth]{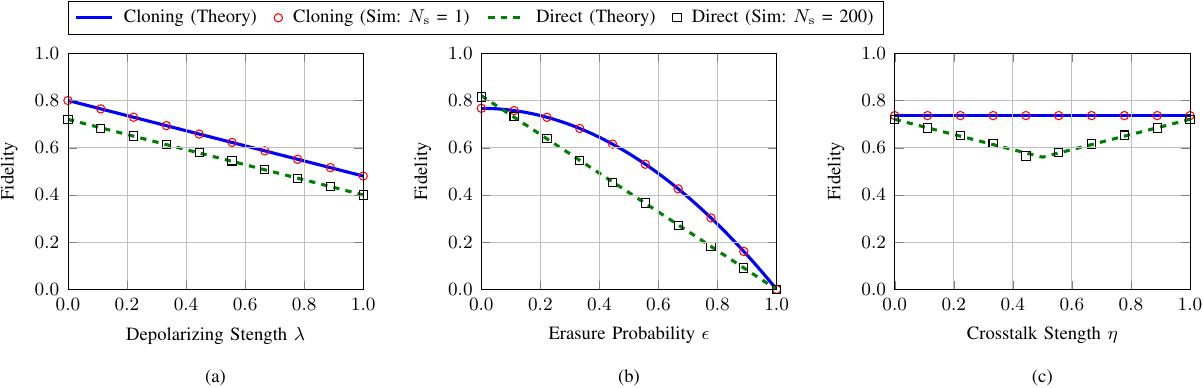}
	
	\caption{Fidelity with multiplexing (Direct) \eqref{eq:multiplexing_fidelity} and with cloning-based diversity (Cloning) \eqref{eq:cloning_fidelity} on a $2\times 2$ \ac{MIMO} channel. The parameters that are not being varied are fixed to 0.2 in each plot.}
	\label{fig:deplineplot0-2}
\end{figure*}
%%%%%%%%%%%%%

Without a loss of generality, we can focus on transmission from \ac{Tx}~1 to the \ac{Rx}~1. The state at the \ac{Rx}~1, for a fixed $\ket{\psi_0}_1$ averaged over all $\ket{\psi}_2$ is
\begin{align}
	\rho_1 &= \ExpO{\tr_2\left\{\QC{D}_{\lambda}^{\otimes 2}\left(
					\QC{E}_{\epsilon}^{\otimes 2}\left( 
					\QC{C}_{\eta}\left(
					\ketbra{\psi}{\psi}_{1, 2}
					\right)
					\right)	\right)\right\}}{\ket{\psi}_2\sim\mathrm{Haar}}\\
				&=\tr_2\left\{\QC{D}_{\lambda}^{\otimes 2}\left(
				\QC{E}_{\epsilon}^{\otimes 2}\left( 
				\QC{C}_{\eta}\left(
				\ExpO{\ketbra{\psi}{\psi}_{1, 2}}{\ket{\psi}_2\sim\mathrm{Haar}}
				\right)
				\right)	\right)\right\}\\
				&= \tr_2\left\{\QC{D}_{\lambda}^{\otimes 2}\left(
				\QC{E}_{\epsilon}^{\otimes 2}\left( 
				\QC{C}_{\eta}\left(
				\ketbra{\psi_0}{\psi_0}_1\otimes\pi
				\right)
				\right)	\right)\right\}.
\end{align}
By utilizing the definitions of all involved maps and evaluating the partial trace, we obtain 
\begin{align}
	\rho_1 &= \left(1 - \epsilon\right)\left[\left(1 - \eta\right)\left(1 - \lambda\right)\ketbra{\psi_0}{\psi_0}_1 \right.+ \nonumber\\
	&~~~~~~~~~~~ \left. \left(1 - \left(1 - \eta\right)\left(1 - \lambda\right)\right) \pi\right] + \epsilon \ketbra{\epsilon}{\epsilon}.
	\label{eq:analysis_1}
\end{align}
This gives the average fidelity
	\begin{align}
		F_{1, 1} &= \bra{\psi_0}\rho_1\ket{\psi_0}\\
					&= \frac{1}{2}\left(1-\epsilon\right)\left[1 +  \left(1 - \eta\right)\left(1 - \lambda\right)\right],
		\label{eq:multiplexing_fidelity}
	\end{align}
which is the same as $F_{2,2}$ due to the same noise processes of equal strengths on both communication channels. Here, we denote by $F_{i, j}$ the fidelity between the $i$th input and the $j$th output of the \ac{MIMO} channels. Finally, noting that the input state $i$ is output at the $i$th output if no swap occurs in the crosstalk. In case of a swap, the input $i$ goes to the $j\neq i$ output port. This observation gives us $F_{1, 2} = F_{2, 1} = \frac{1}{2}\left(1-\epsilon\right)\left[1 +  \eta\left(1 - \lambda\right)\right]$.

\subsubsection{Diversity}
In diversity, a single qubit is transmitted over a single use of the multiple \ac{MIMO} channels. In classical communication systems, one typically transmits multiple copies of the same signal over multiple antennas (diversity) to improve the communication reliability. Due to the reception of multiple copies of the same signal---where each copy is degraded differently due to stochasticity---the receiver is able to reconstruct a higher quality signal by \emph{combining} all received signals. Following the same rationale, we feed a copy of the qubit to be transmitted in each input port in hopes to improve the communication reliability, i.e., improve the fidelity of the state at \ac{Rx} with respect to the original information. However, due to no-cloning theorem, we cannot prepare perfect copies of the qubit. We resort to the \acp{UQCM} to prepare approximate copies of the qubit to be transmitted and feed these approximate copies in the each input port \cite{BL:19:Book, KW:99:JMP, BEM:98:PRL, Wer:98:PRA, SIG:05:RMP}. 

Let $\ket{\psi}$ be the qubit to be transmitted. The transmitter uses the \ac{UQCM} to obtain
\begin{align}
	\sigma^{\mathrm{clone}}_{1, 2} &= \frac{2}{3}\ketbra{\psi}{\psi}\otimes\ketbra{\psi}{\psi} + \nonumber\\
	&\frac{1}{6}\left(\ket{\psi}\ket{\psi^{\perp}} + \ket{\psi^{\perp}}\ket{\psi}\right)\left(\bra{\psi}\bra{\psi^{\perp}} + \bra{\psi^{\perp}}\bra{\psi}\right),
	\label{eq:clones}
\end{align}
where $\ket{\psi^{\perp}}$ is the state perpendicular to $\ket{\psi}$. 
First note that $\sigma_1 = \tr_2\left\{ \sigma^{\mathrm{clone}}_{1, 2}\right\} = \frac{5}{6}\ketbra{\psi}{\psi} +\frac{1}{6}\ketbra{\psi^{\perp}}{\psi^{\perp}} = \sigma_2$, giving the fidelity 5/6, even before the transmission. However, soon we will see that this strategy of  predegradation in fidelity still ends up giving a better fidelity in some scenarios. 

For analysis, first note that the \emph{clones are symmetric and thus remain unaffected from the crosstalk}.  Before continuing further, we describe the trivial processing at the receiver to simplify our analysis. At the receiver, no joint processing of the clones is performed. If one of the output ports emits a quantum state without erasure, it is passed to the receiver. If both of the clones are received (no erasure event occurred), one of the clones is discarded and the other one is relayed to the receiver. This allows us to focus on the case when at least one clone is received and we can write the output state of any of the two clones in \eqref{eq:clones}. At least one of the clones is received with probability $1 - \epsilon^2$ and its state is
\begin{align}
		\rho_1^{\mathrm{clone}} &=
					 \tr_2\left\{\QC{D}_{\lambda}\otimes\QC{D}_{\lambda}\left(
					\QC{E}_{\epsilon}\otimes\QC{E}_{\epsilon}\left( 
					\QC{C}_{\eta}\left(
					\sigma^{\mathrm{clone}}_{1, 2}
					\right)
					\right)	\right)\right\}\\
					&= \frac{5}{6}\left(1 - \lambda\right) \ketbra{\psi}{\psi} + \frac{1}{6}\left(1 - \lambda\right) \ketbra{\psi^{\perp}}{\psi^{\perp}} + \lambda\, I/2,
					\label{eq:analysis_2}
\end{align}
giving the average output fidelity 
\begin{align}
	F_{1, 1}^{\mathrm{clone}} = \left(1 - \epsilon^2\right)\left(\frac{5}{6} - \frac{1}{3}\lambda\right).
	\label{eq:cloning_fidelity}
\end{align}
Due to the symmetry of the overall system including the inputs, we have $F_{i, j}^{\mathrm{clone}} = F_{1, 1}^{\mathrm{clone}}$, for all $i, j \in \left\{1, 2\right\}$.

Fig.~\ref{fig:deplineplot0-2} shows the comparison of achieved fidelity with direct transmission of two data streams on a $2\times 2$ \ac{MIMO} setup (multiplexing) and the cloning-based diversity scheme. We plot the two fidelity expressions obtained above for theoretical results and simulate the system with Haar random quantum states. The effective impairments after cloning in this setup are only the losses (erasure) and the depolarizing noise, both of which are isotropic. Thus, the simulation does not require multiple random samples for convergence in this case. Therefore, we use $N_{\mathrm{s}} = 1$ samples in simulating the diversity approach but set $N_{\mathrm{s}} = 200$ for the direct multiplexing approach, where multiple samples are required for the crosstalked state to converge to the maximally mixed state. From these numerical results, we see a tradeoff between the multiplexing (we can transmit 2 qubits per \ac{MIMO} use albeit with lower average fidelity) and diversity (1 qubit per \ac{MIMO} use but a higher average fidelity) in specific regions of the parameter values. 

\begin{figure}[t!]
	\centering
	\includegraphics[width = 0.45\textwidth]{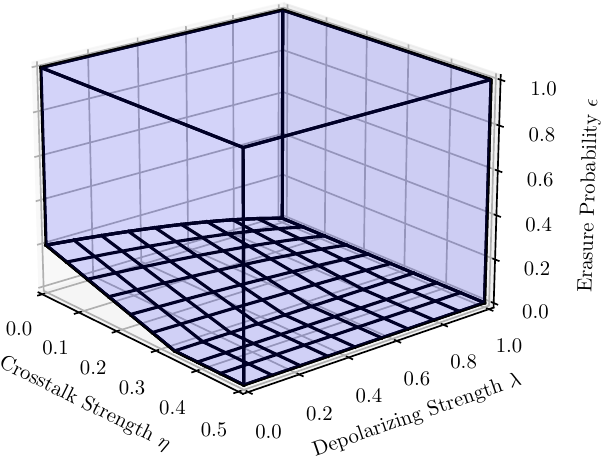}
	\caption{ Region of the diversity gain. The box shows the boundary of $\left(\eta, \epsilon, \lambda\right)$-region inside of which \ac{SQC} provides a diversity gain.}
	\label{fig:region}
\end{figure}

Fig.~\ref{fig:region} shows the region in $\left(\eta, \epsilon, \lambda\right)$-space for $0\leq \eta \leq 0.5$ and $0 \leq \epsilon, \lambda \leq 1$ where the cloning based diversity approach provides a diversity gain in fidelity. That is, \eqref{eq:cloning_fidelity} has a higher value than \eqref{eq:multiplexing_fidelity}. This region of diversity gain covers the $\approx 95\%$ of the valid parameter region region.\footnote{Based on evaluating both fidelity expressions on a uniform grid of $8.0\times 10^6$ points.}

\section{Generalization to $2^m \times 2^m$ MIMO Setup}
In the absence of crosstalk impairments, generalizing the system model of Fig.~\ref{fig:diversitycrosstalk} to $N\times N$ \ac{MIMO} is straightforward since the remaining impairments---losses and noise---are independent of each other and considered symmetric on all channels. Generalizing the system model in the presence of crosstalk can be challenging. 

\begin{figure}[t!]
	\centering
		\includegraphics[width=0.35\textwidth]{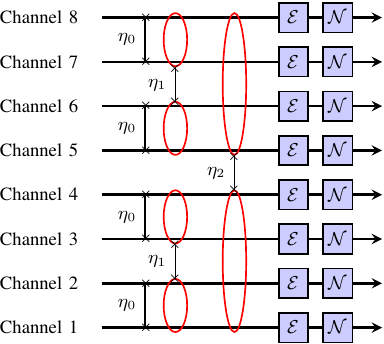}
	\caption{Generalization to $2^3\times 2^3$ \ac{MIMO} channel. The crosstalk model of Fig.~\ref{fig:diversitycrosstalk}(b) is applied recursively. } 
	\label{fig:diversitygeneral}
\end{figure}

Here, we adopt our crosstalk model recursiverly to model $N\times N$ quantum \ac{MIMO} setup for $N = 2^m$. Our approach is exemplified in Fig.~\ref{fig:diversitygeneral} for $m = 3$. The crosstalk occurs in $m$ steps. In the $i$th step $2^m$ channels are divided into $2^{m - i}$ groups of $2^i$ channels each for $i \in \left\{0, \cdots, m - 1\right\}$. In each step, one group of channel crosstalks with one and exactly one other group of the channels as a whole. The grouping can be motivated by the spatial structure of the channels; two channels close to each other have higher chance of crosstalk with each other and thus should be made to crosstalk with each other directly at the first stage. Two groups that crosstalked in the $i$th step will be grouped together to form a single group in the next step. Crosstalk at $i$th step is parameterized by $\eta_i$. We assume $\eta_i > \eta_{i+1}$ to reflect that the channels grouped in earlier stages of crosstalk modeling are closer in proximity and hence induce stronger crosstalk. 

Again, we utilize $1 \rightarrow 2^x$ \ac{SQC} for $x\leq m$ for achieving diversity in $2^{m - x}$ independent data streams. The $2^x$ clones are fed into the group of channels that crosstalk outside of the group as late as possible. For example, for $x = 1$ gives two clones each of independent data streams. Clones of the same data qubit can be fed into channel group $\left\{1, 2\right\}$, but not to the group $\left\{2, 3\right\}$ because the former group has not crosstalk outside of the group in the first layer of crosstalk but the later does. With this setting, we can apply the same ideas from \eqref{eq:analysis_1} and \eqref{eq:analysis_2} recursively to obtain the fidelity 
\begin{align}
	F_{1, 1}^{\left(m, x\right)} = \left(1 - \epsilon^{2^x}\right)\left[ X\, F_{1 \rightarrow 2^x} + \frac{1}{2} \left( 1 - X\right)\right],
	\label{eq:fidelity_mx}
\end{align}
where 
$$F_{1 \rightarrow M} = \frac{2M + 1}{3M}$$
 is the fidelity of each clone after $1\rightarrow M$ \ac{SQC} \cite{SIG:05:RMP} and 
 $$
 	X = \left(1 - \lambda\right)\prod_{i=x}^{m-1}\left(1 - \eta_{i}\right).
 $$

\begin{figure}[t!]
	\centering
	\includegraphics[width=0.475\textwidth]{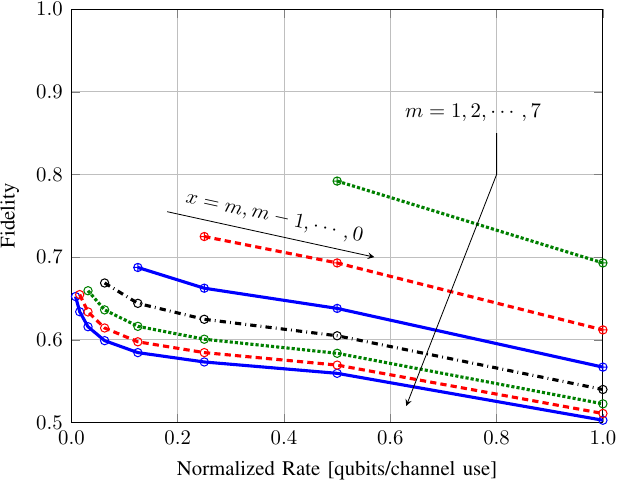}
	\caption{Achievable \ac{DMT} with cloning-based approach in the general \ac{MIMO} channels with crosstalk, losses, and depolarizing noise.}
	\label{fig:dmt-1-crop}
\end{figure}

From \eqref{eq:fidelity_mx}, we can plot an achievable \ac{DMT} with our proposed scheme for quantum \ac{MIMO} channel. Analyzing \eqref{eq:fidelity_mx} we see that the fidelity does increase by increasing the diversity order in moderately or highly noisy channels. This tradeoff is shown in Fig.~\ref{fig:dmt-1-crop} with $\epsilon = \lambda = 0.1$ and $\eta_i = 0.4/(1.2)^i$ for $m = 1, 2, \cdots, 7$.

However, consider the case when the \ac{MIMO} setup has a very low noise, e.g., consider the noiseless case $\epsilon \approx \lambda \approx \eta_i \approx 0$ for all $i$. In this setting, $F_{1, 1}^{\left( m, x \right)} = F_{1 \rightarrow 2^x}$, which decreases by increasing $x$. That is, the fidelity decreases by increasing the diversity order. This performance degradation arises from the (lack of) information combining at the receiver, whereas the process of approximate cloning results in reduced fidelity even before the propagation impairments. To avoid this, one can either (a) use feedback from the receiver to fix $x$ at a favorable value, or (b) develop a novel information combining technique at the receiver that can distill a high-fidelity copy of the target state from multiple received clones. Whether a method achieving (b) exists remains an open question. 

The process of $1 \rightarrow M$ approximate quantum cloning is a unitary operation on $M$ principle qubits (including the input state) and some ancilla qubits. In the case of availability of all $M$ clones and the ancilla qubits, the information concentration is trivial (unitary inversion) to obtain the original noiseless qubit back. Even when the clones and ancilla qubits are not locally available, remote information concentration is possible with local operations and classical communication \cite{MV:01:PRL, WZY:11:PRA}. However, these solutions do not cover the case that we consider here since (i) the ancilla is not transmitted on the channel, (ii) some of the clones are lost, and (iii) all clones have undergone noisy evolution. The possibility of information concentration when only $k$ of $M$ clones are available is ambiguous.

\section{Conclusions}
\ac{MIMO} techniques in classical communications brought a paradigm shift, which makes the latest generations of digital communication possible. Here, we have shown that quantum \ac{MIMO} communication also has a potential that can improve the communication reliability and communication rates. In this initial study on the topic, we have demonstrated a tradeoff between the diversity and multiplexing when multiple transmit and receive apertures are employed. In our simple communication strategy, no information combining is performed at the receiver and hence our results can be considered as the minimum achievable performance. In future, it will be interesting to see the development of information combining techniques based on our proposed processing at the \ac{Tx} or the development of altogether new techniques with improved performance, fully exploiting the \ac{MIMO} potential of quantum communications. Furthermore, quantum communication channels are known to exhibit counter-intuitive behavior in several situations when multiple channels are combined, e.g., superadditivity of capacities and nonconvexity. Exploring these effects in quantum \ac{MIMO} communications can be an interesting direction of research.



\begin{thebibliography}{10}
	\providecommand{\url}[1]{#1}
	\csname url@samestyle\endcsname
	\providecommand{\newblock}{\relax}
	\providecommand{\bibinfo}[2]{#2}
	\providecommand{\BIBentrySTDinterwordspacing}{\spaceskip=0pt\relax}
	\providecommand{\BIBentryALTinterwordstretchfactor}{4}
	\providecommand{\BIBentryALTinterwordspacing}{\spaceskip=\fontdimen2\font plus
		\BIBentryALTinterwordstretchfactor\fontdimen3\font minus
		\fontdimen4\font\relax}
	\providecommand{\BIBforeignlanguage}[2]{{%
			\expandafter\ifx\csname l@#1\endcsname\relax
			\typeout{** WARNING: IEEEtran.bst: No hyphenation pattern has been}%
			\typeout{** loaded for the language `#1'. Using the pattern for}%
			\typeout{** the default language instead.}%
			\else
			\language=\csname l@#1\endcsname
			\fi
			#2}}
	\providecommand{\BIBdecl}{\relax}
	\BIBdecl
	
	\bibitem{Wit:91:conf}
	A.~Wittneben, ``Basestation modulation diversity for digital simulcast,'' in
	\emph{{IEEE} 41st Vehicular Technology Conference}, May 1991, pp. 848--853.
	
	\bibitem{SW:93:conf}
	N.~Seshadri and J.~Winters, ``Two signaling schemes for improving the error
	performance of frequency-division-duplex ({FDD}) transmission systems using
	transmitter antenna diversity,'' in \emph{{IEEE} 43rd Vehicular Technology
		Conference}, May 1993, pp. 508--511.
	
	\bibitem{Ala:98:JSAC}
	S.~Alamouti, ``A simple transmit diversity technique for wireless
	communications,'' \emph{IEEE Journal on Selected Areas in Communications},
	vol.~16, no.~8, pp. 1451--1458, Oct. 1998.
	
	\bibitem{ZT:03:TIT}
	L.~Zheng and D.~Tse, ``Diversity and multiplexing: a fundamental tradeoff in
	multiple-antenna channels,'' \emph{IEEE Transactions on Information Theory},
	vol.~49, no.~5, pp. 1073--1096, May 2003.
	
	\bibitem{SC:02:conf}
	E.~Shin and V.~Chan, ``Optical communication over the turbulent atmospheric
	channel using spatial diversity,'' in \emph{{IEEE} Global Telecommunications
		Conference, 2002. {GLOBECOM} '02}, vol.~3, Nov. 2002, pp. 2055--2060.
	
	\bibitem{RS:05:TWC}
	M.~Razavi and J.~Shapiro, ``Wireless optical communications via diversity
	reception and optical preamplification,'' \emph{IEEE Transactions on Wireless
		Communications}, vol.~4, no.~3, pp. 975--983, May 2005.
	
	\bibitem{JB:19:TIT}
	A.~Jaiswal and M.~R. Bhatnagar, ``Free-space optical communication: A
	diversity-multiplexing tradeoff perspective,'' \emph{IEEE Transactions on
		Information Theory}, vol.~65, no.~2, pp. 1113--1125, Feb. 2019.
	
	\bibitem{KCS:22:CST}
	S.~Koudia, A.~S. Cacciapuoti, K.~Simonov, and M.~Caleffi, ``How deep the theory
	of quantum communications goes: Superadditivity, superactivation and causal
	activation,'' \emph{IEEE Communications Surveys \& Tutorials}, vol.~24,
	no.~4, pp. 1926--1956, 2022.
	
	\bibitem{WZ:82:Nat}
	W.~K. Wootters and W.~H. Zurek, ``A single quantum cannot be cloned,''
	\emph{Nature}, vol. 299, no. 5886, pp. 802--803, Oct. 1982.
	
	\bibitem{Som:20:PRA}
	S.~Bandyopadhyay, ``Impossibility of creating a superposition of unknown
	quantum states,'' \emph{Phys. Rev. A}, vol. 102, p. 050202, Nov 2020.
	
	\bibitem{OGH:16:PRL}
	M.~Oszmaniec, A.~Grudka, M.~Horodecki, and A.~W\'ojcik, ``Creating a
	superposition of unknown quantum states,'' \emph{Phys. Rev. Lett.}, vol. 116,
	p. 110403, Mar 2016.
	
	\bibitem{YC:20:TCOM}
	R.~Yuan and J.~Cheng, ``Free-space optical quantum communications in turbulent
	channels with receiver diversity,'' \emph{IEEE Transactions on
		Communications}, vol.~68, no.~9, pp. 5706--5717, Sep. 2020.
	
	\bibitem{WRA:24:arX}
	Z.~Wang, T.~C. Ralph, R.~Aguinaldo, and R.~Malaney, ``Exploiting spatial
	diversity in earth-to-satellite quantum-classical communications,''
	\emph{arXiv preprint arXiv: 2407.02224}, Jul. 2024.
	
	\bibitem{KDM:21:CL}
	N.~K. Kundu, S.~P. Dash, M.~R. McKay, and R.~K. Mallik, ``{MIMO} {Terahertz}
	quantum key distribution,'' \emph{IEEE Communications Letters}, vol.~25,
	no.~10, pp. 3345--3349, Oct. 2021.
	
	\bibitem{KMC:23:TQE}
	N.~K. Kundu, M.~R. McKay, A.~Conti, R.~K. Mallik, and M.~Z. Win, ``{MIMO}
	{Terahertz} quantum key distribution under restricted eavesdropping,''
	\emph{IEEE Transactions on Quantum Engineering}, vol.~4, pp. 1--15, 2023.
	
	\bibitem{ZPD:23:TQE}
	M.~Zhang, S.~Pirandola, and K.~Delfanazari, ``Millimeter-{Waves} to {Terahertz}
	{SISO} and {MIMO} continuous variable quantum keydistribution,'' \emph{IEEE
		Transactions on Quantum Engineering}, vol.~4, pp. 1--10, 2023.
	
	\bibitem{KOB:24:arXiv}
	S.~Koudia, L.~Oleynik, M.~Bayraktar, J.~ur~Rehman, and S.~Chatzinotas,
	``Physical layer aspects of quantum communications: A survey,'' \emph{arXiv
		preprint arXiv: 2407.09244}, Jul. 2024.
	
	\bibitem{BL:19:Book}
	D.~Bruss and G.~Leuchs, Eds., \emph{Quantum information: from foundations to
		quantum technology applications}, second edition~ed.\hskip 1em plus 0.5em
	minus 0.4em\relax Weinheim, Germany: Wiley-VCH, 2019.
	
	\bibitem{KW:99:JMP}
	M.~Keyl and R.~F. Werner, ``Optimal cloning of pure states, testing single
	clones,'' \emph{J. Math. Phys.}, vol.~40, pp. 3283--3299, Jul. 1999.
	
	\bibitem{BEM:98:PRL}
	D.~Bruss, A.~Ekert, and C.~Macchiavello, ``Optimal universal quantum cloning
	and state estimation,'' \emph{Phys. Rev. Lett.}, vol.~81, no.~12, pp.
	2598--2601, Sep. 1998.
	
	\bibitem{Wer:98:PRA}
	R.~F. Werner, ``Optimal cloning of pure states,'' \emph{Physical Review A},
	vol.~58, no.~3, pp. 1827--1832, Sep. 1998.
	
	\bibitem{SIG:05:RMP}
	V.~Scarani, S.~Iblisdir, N.~Gisin, and A.~Acín, ``Quantum cloning,''
	\emph{Reviews of Modern Physics}, vol.~77, no.~4, pp. 1225--1256, Nov. 2005.
	
	\bibitem{Note1}
	Based on evaluating both fidelity expressions on a uniform grid of $8.0\times
	10^6$ points.
	
	\bibitem{MV:01:PRL}
	M.~Murao and V.~Vedral, ``Remote information concentration using a bound
	entangled state,'' \emph{Physical Review Letters}, vol.~86, no.~2, pp.
	352--355, Jan. 2001.
	
	\bibitem{WZY:11:PRA}
	X.-W. Wang, D.-Y. Zhang, G.-J. Yang, S.-Q. Tang, and L.-J. Xie, ``Remote
	information concentration and multipartite entanglement in multilevel
	systems,'' \emph{Phys. Rev. A}, vol.~84, p. 042310, Oct 2011.
	
\end{thebibliography}
\end{document}